\documentclass{article}
\usepackage{frascatiphys}
\usepackage{graphicx}
\begin{document}
\title{ 
The Milky Way Hot Baryons and their Peculiar Density Distribution: a Relic of Nuclear Activity}
\author{
Fabrizio Nicastro$^{1,2}$, Francesca Senatore$^{1}$, Yair Krongold$^{3}$, Smita Mathur$^{4}$, Martin Elvis$^{2}$ \\
$^1$ {\em INAF - Osservatorio Astronomico di Roma, Via Frascati, 33, 00078 Monte Porzio Catone (RM), Italy} \\
$^2$ {\em Harvard-Smithsonian Center for Astrophysics, Cambridge, MA, USA} \\
$^3$ {\em Instituto de Astronomia - UNAM, Mexico City (DF), Mexico} \\
$^4$ {\em Ohio State University,  Columbus OH, USA} \\}
\maketitle
\baselineskip=11.6pt
\begin{abstract}
We know that our Galaxy is permeated by tenuous, hot, metal-rich gas. However much remains unknown about its origin, the portion of the Galaxy that it permeates, 
its total mass, as any role it may play in regulating activity in the Galaxy. 
In a Letter currently in the press with the ApJ, we show that this hot gas permeates both the disk of the Galaxy and a large spherical volume, centered on the Galactic 
nucleus, and extending out to distances of at least 60-200 kpc from the center. This gas displays a peculiar density distribution that peaks about 6 kpc from the Galaxy's 
center, likely witnessing a period of strong activity of the central supermassive black hole of the Milky Way that occurred 6 Myrs ago. With our study we are also able to 
update the total baryonic mass of the Galaxy to M$_b = (0.8-4.0) \times 10^{11}$  M$_{\odot}$, sufficient to close the Galaxy's baryon census. 
\end{abstract}
\baselineskip=14pt

\section{Introduction}
The visible baryonic mass of the Milky Way amounts to M$_b^{Obs} \simeq 0.65×\times 10^{11}$  M$_{\odot}$ $^{1)}$. 
The total baryonic plus dark matter mass of our Galaxy, is instead M$_{Tot} \simeq (1-2) \times 10^{12}$  M$_{\odot}$  $^{(2)}$. 
This, assuming a universal baryon fraction of $f_b = 0.157$ $^{(3)}$, implies a total baryonic mass of M$_b^{Pred} \simeq (1.6-3.2) \times 10^{11}$  
M$_{\odot}$, between 2.5 and 5 times larger than observed. A large fraction of the baryonic mass of our Galaxy is thus currently eluding detection. 

This “missing-baryon” problem is not a monopoly of the Milky Way: most of the galaxies in the local universe suffer a deficit of baryonic mass compared to their dynamical 
mass and the problem is more serious at smaller dynamical masses (e.g. (4)), suggesting that lighter galaxies fail to retain larger fractions of their baryons. 
These baryons could be at least partly hiding under the form of tenuous hot ($\sim 10^6$ K) gas, heated up by recurring episodes of nuclear activity during the galaxy's 
lifetime, such as bursts of star formation followed by powerful supernova explosions or accretion-powered ignitions of the central supermassive black hole, which may 
have powered energetic outflows that pushed material out to large distances from the Galaxy's center. 

Over the past several years, a number of experiments, as well as theoretical works, have attempted to gain insights into the location and mass of the 
hot medium in our own Galaxy. 
Our peripheral position in the Galaxy, at about 8.5 kpc from the Galaxy's center and roughly in the Galaxy's plane, gives us hope of solving the problem: once 
a physically motivated density profile is assumed for the hot absorbing medium, the observed column densities (as well as the other observables) will depend critically 
on the sky position (and distance, for Galactic background targets) of the sources towards which the column densities are measured. 
This consideration has recently motivated several studies, which have used available spectra of extragalactic targets, with no other selection 
criterion than being at high Galactic latitudes, to measure OVII column densities and compare them with physically motivated or simple phenomenological density profile 
models $^{(5,6,7)}$. 
The results, however, are often contradictory, with estimated total masses of the million degree medium within a 1.2 virial-radius sphere (300 kpc) 
that strongly depend on the flatness of the assumed density profile, and range from a negligible M$_{Hot} \simeq 2.4 \times 10^9$  M$_{\odot}$ $^{(5)}$ to 
a significant M$_{Hot} \simeq 10^{11}$ M$_{\odot}$ $^{(7)}$.  

\noindent
Here we present an experiment that settles the controversy by adopting a number of novel and rigourous data analysis techniques and sample selection criteria, 
and that has recently been published by the ApJ Letter ($^{(8)}$, hereinafter N16). 

\noindent
Throughout this contribution, we refer to all densities and masses in units of ($A_O/4.9\times 10^{-4})^{-1} \times [Z/(0.3 Z_{\odot})]^{-1} (f_{OVII}/0.5)^{-1}$, where $A_O$ is 
the relative abundance of oxygen compared to hydrogen, $Z$ is the metallicity and $f_{OVII}$ is the fraction of OVII relative to oxygen. 
For easy comparison to other works (e.g. $^{(7)}$), we compute hot baryon masses within a 1.2 virial-radius sphere. 
Errors on best-fitting parameters (and quantities derived from those) are provided at 90\% confidence level for a number of interesting parameters equal to 
(31-N$_{dof}$), where N$_{dof}$ is the number of degrees of freedom in the fit.

\section{Data, Model Components and Procedures}

\subsection{LGL and HGL Samples}
Our LGL+HGL sample differs from those previously used to perform analyses similar to ours (e.g.$^{(5,9)}$) in three important ways: 
(1) for the first time we use simultaneously HGL and LGL samples;  
(2) our two XMM-{\em Newton} Reflection Grating Spectrometer (RGS) samples are complete to a minimum Signal to Noise per Resolution Element in the continuum, 
SNRE$>$10 at 22 \AA; 
(3) whenever possible (see below), we  remove the degeneracy between column density and Doppler parameter of the instrumentally unresolved 
OVII lines, by performing a detailed curve of growth analysis (e.g. $^{(10)}$). 

Our final HGL and LGL samples contain 18 and 13 lines of sight, respectively, leading to a well-defined SNRE-complete observed distribution of 31 OVII K$\alpha$ 
EWs and sky positions (see Figure 1 in N16). 
 
HLG OVII absorbers spread over more than an order of magnitude in column densities, from a minimum value of N$_{OVII} = 0.8_{-0.5}^{+1.2} \times 10^{16}$ cm$^{-2}$, to a 
maximum value of N$_{OVII} = 33_{-29}^{+480} \times 10^{16}$ cm$^{-2}$. The spread is less extreme for LGL absorbers that span a factor of about 5 in OVII column densities. 

\subsection{Functions and Fitting Procedure}
We model the derived distribution of 31 column densities and sky positions with two most general families of density profiles, i.e.: spherically-symmetric (SS), where 
the only scale-length parameter is the core-radius $R_c$ (exponential- and $\beta$-profile models), and Cylindrically-Symmetric (CS), characterized by two different 
scale-length parameters, the coplanar core-radius $\rho_c$ and the vertical core-height $h_c$ (again, exponential- and $\beta$-profile models: see N16 for the analytical 
expressions adopted). 
For each functional form, we also allow for the inclusion of an additional parameter ($R_s$ for the SS profiles and $h_s$ for the CS profiles, both in kpc) allowing for a 
possible offset of the distributions from the Galaxy's center (SS models) or plane (CS models).  

\subsection{Halo Extent and Masses}
Given the one-dimensional nature of our observables, only a lower limit to the total extent of the volume containing the hot absorbing gas seen against HGL targets 
can be evaluated in our analyses. We evaluate this limit by stopping the line of sight integration at a line of sight distance $\xi$ where the relative difference between 
two consecutive values of the column density differs by less than 0.01\% (much lower than the typical relative uncertainty on our column density measurements, which 
is of the order of $\simeq 10$\% in the best cases). Under the assumption of a centrally symmetric halo, the largest of these line of sight distances in the best-fitting 
HGL models, sets effectively a lower limit to the radial size of the halo, and so its baryon mass. Smaller halos are not allowed by the necessity to accumulate sufficient 
column density (and emission measure) along the thickest HGL lines of sight. On the other hand, larger halo sizes (and therefore baryon masses) are clearly possible, 
but not directly measurable through our observables. 

\subsection{Caveats on Parameter Degeneracy}
For the same limitation intrinsic in the one-dimensionality of our observables, the parameters of our models are all degenerate, to some extent. 
For exponential profiles, the problem is negligible and only a moderate degeneracy is present between the peak density $n_0$ and the scale-distance parameters 
$R_c$ or $\rho_c$ and $h_c$. 
For $\beta$-like profiles, instead, the scale-distance parameters are often strongly degenerate with the flatness index $\beta$ and, when this happens, is impossible to 
discriminate statistically between very steep and compact (exponential) or flat and extended profiles. 
In these cases exponential profiles or $\beta$-like profiles with either the flatness index $\beta$ or radial scale distance $R_c$ frozen to some physically motivated 
value, provides the only non-degenerate solutions (see N16 Tables and details). 
In all cases, however, the simultaneous modeling of LGLs and HGLs, together with the presence of a radial offset in the distribution (i.e. best-fitting $R_s > 0$), tend to break 
the degeneracy between scale-distance and the index $\beta$. When this happens, the shape of the density profile can be determined and all model 
parameters are generally robustly constrained to physically reasonable best-fitting values (models A and B in table 2 of N16), and so are the derived minimum extent 
and mass of the halo. 

\section{Modeling and Results}
First we modeled the HGL and LGL separately. The 18 HGL absorbers are equally well fitted by both SS and CS models, and the additional degree of complexity 
introduced by CS models over SS models is not statistically required. Density profiles are generally steep and, interestingly, the best-fitting profile is exponential 
and has an offset-radius $R_s =5.4_{-0.4}^{+0.6} $ kpc (see details in N16). This shift in radius would indicate that the hot baryon density in the halo increases radially 
from the Galaxy's center up to its peak value at 5.4 kpc, and then decreases monotonically towards the virial radius. 
The implied baryonic mass is unimportant $M_{Hot}^{Halo} =3.3_{-1.4}^{+4.1}  \times 10^9$ M$_{\odot}$ (in full agreement with the mass derivable from the best-fitting 
parameters of the “spherical-saturated” model of $^{(5)}$: see their table 2). 

Unlike the HGL absorbers, for the 13 members of our LGL sample a flattened disk-like CS density profile is statistically greatly preferred to a SS profile [$\chi_{Flat}^2 = 12.8$ 
for 10 degrees of freedom (dof), versus $\chi_{Sph}^2(dof) = 22.5(11)$]. The LGL absorbers are clearly tracing a disk-like distribution in the Galaxy's plane, with best-fitting 
radial and height scale lengths in excellent agreement with those of the stellar disk of the Milky Way  $^{(11)}$ (see details in N16). 
The mass of the hot gas in the Galactic disk is only $M_{Hot}^{Disk} = 1.4_{-0.6}^{+1.1}  \times 10^8$ M$_{\odot}$, of the order of that of the other gaseous components of the 
disk $^{(11)}$. 

The models that best-fit separately our HGL (halo) and LGL (disk) absorbers, are very different in both their central densities and profiles (Tab. 1 in N16), and neither of the 
two can adequately model the column-density distribution of the other. 
Either a compromise single-component model is needed, or the two components must be physically distinct

We then proceeded to model simultaneously and self-consistently all 31 HGL and LGL lines of sight. 
We tried two alternative families of functions: (A) a single-component set of models, with all parameters free to vary and the normalizing density peaking at 
$R = R_s$ (A-type models, hereinafter), and (B) a 2-component set of models in which a parameter-varying SS component with $n=0$ for $R<R_s$ and 
$n = n(R)$ for $R \ge R_s$, is added to a flattened disk-component with parameters frozen to the LGL best-fitting values (B-type models, hereinafter). 
Both sets of models can provide statistically acceptable fits (see Table 2 in N16). In both cases offset radii $R_s > 0$ are statistically preferred 
(compare A with M4 and B with M3 in Fig. 1a,b, and see N16 for details). 

Our two best-fitting models A and B have offset radii $R_s =5.6_{-0.6}^{+0.6}$  kpc and $R_s = 6.7_{-1.8}^{+0.9}$ kpc, consistent with each other and 
with the best-fitting value found by fitting the HGL sample only. 
In model A, $R_s =0$ is ruled out at a 4-interesting-parameter statistical significance of 14.9$\sigma$. Similarly, for our alternative best-fitting model B, 
$R_s = 0$ is excluded at a 4-interesting-parameter statistical significance of 6.0$\sigma$. 
 
Fig. 1a and 1b show that Model A (left 2 panels of Fig. 1a) is able to better reproduce the observed spread of OVII column densities along HGL lines of sight, compared to 
model B (left 2 panels of Fig. 1b), which however (by construction) reproduces better the observed LGL columns (see N16 for details). By comparison, the alternative two 
models with $R_s$ frozen to zero, M3 and M4, clearly reproduce HGL and LGL columns less well than the respective models B and A. 
The actual solution lies probably in between models A and B. 
\begin{figure}[htb]
    \begin{center}
        {\includegraphics[scale=0.27]{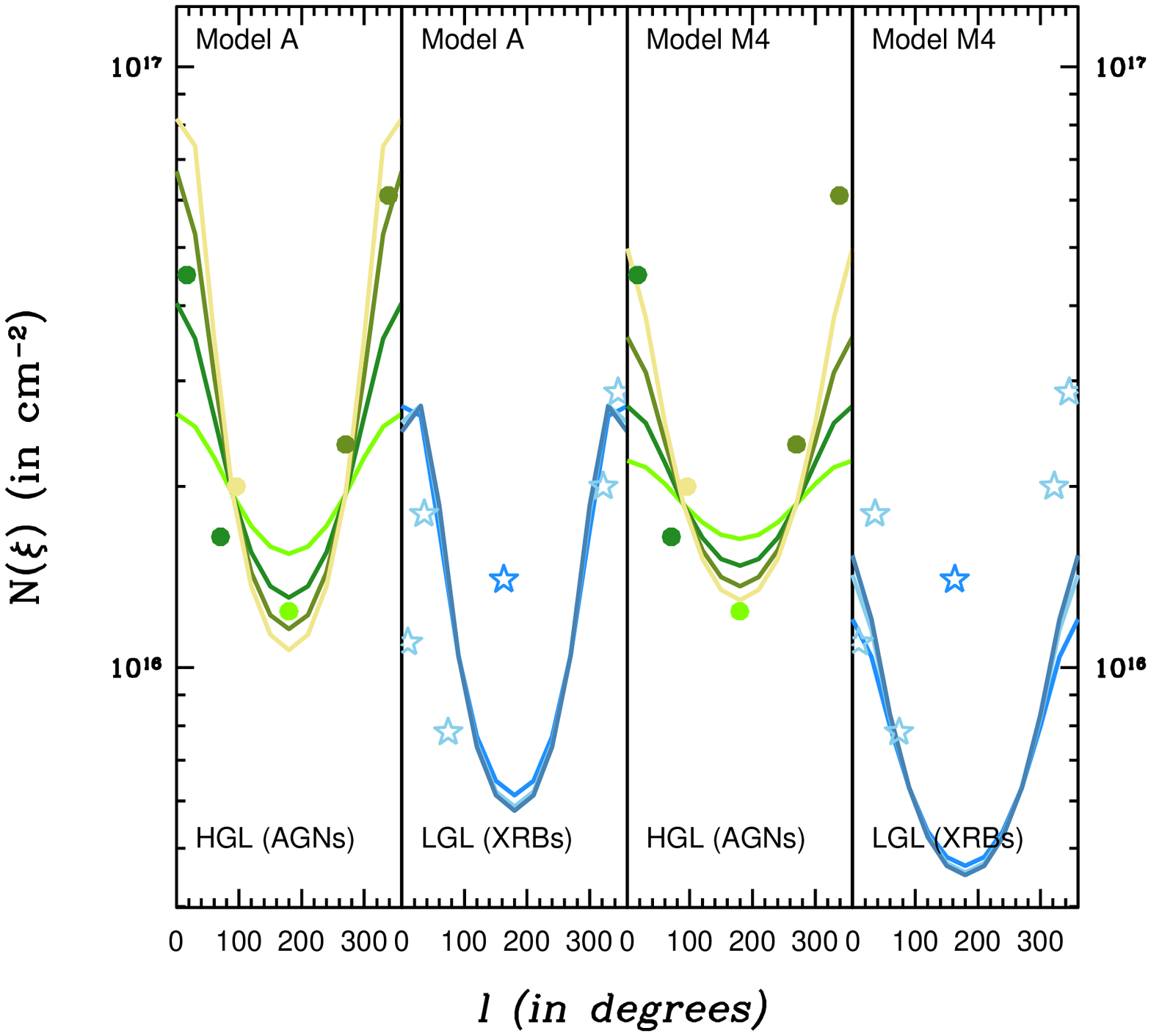}}\hspace{0.5cm}
        {\includegraphics[scale=0.27]{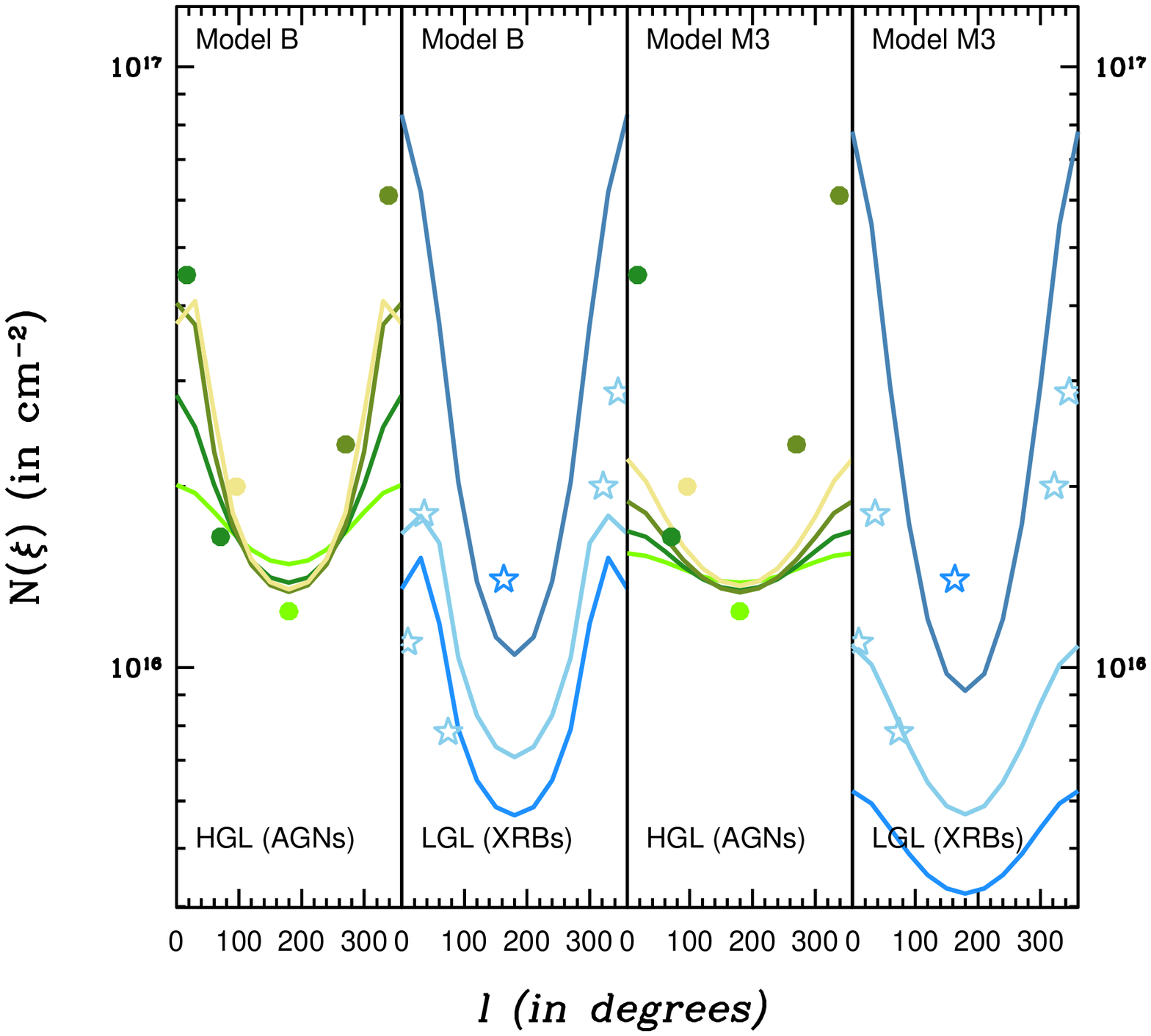}}
        \caption{\it {\bf Figure 1a} (left): Distributions of column densities versus Galactic longitude (green and blue curves) for different values of Galactic latitudes 
(different color gradation), predicted by our best- fitting models A (left 2 panels), compared to the same quantities predicted by the corresponding models with 
$R_s$ frozen to zero, M4 (2 right panels). 
Filled green circles and blue stars are our HGL and LGL data, binned in $\Delta l=30^0$ bins of Galactic longitudes, and the color gradation corresponds to different 
values of Galactic latitude. For HGLs: $15^0<|b|<75^0$ from dark to light green. For LGLs: $|b|<30^0 $from light to dark blue. 
{\bf Figure 1b} (right): same as Figure 1a, but for our best-fitting model B (left 2 panels), compared to the corresponding model with $R_s$ frozen to zero, M3 (2 right panels)}
\label{fig23}
    \end{center}
\end{figure}

Our best-fitting models A and, to a lesser extent, B also reproduce well the Emission Measure at high Galactic latitudes and towards the Galactic center (see N16 for 
details). 

From our best-fitting models we derive total hot baryon masses in the ranges M$_{Hot}(A) =0.2_{-0.1}^{+0.3}  \times 10^{11}$ M$_{\odot}$ and M$_{Hot}(B) = 1.3_{-0.7}^{+2.1}  \times 
10^{11}$ M$_{\odot}$. These masses are $>10$ times larger than those obtained by fitting the HGL sample only. This is due to the flatness of the best-fitting 
density profiles (see N16 for details). 
These flat profiles imply minimum sizes of the halo of $>60$ kpc and $>200$ kpc for Model A and B, respectively. 

Adding the hot baryon mass to the visible mass of the Milky Way, gives a total baryonic mass in the range M$_b = (0.8-4.0) \times 10^{11}$ M$_{\odot}$, 
sufficient to close the Galaxy's baryon census.
 
\subsection{Discussion}
Our analysis indicates not only (1) that both the Galactic plane and the halo are permeated by OVII-traced million degree gas, but also (2) that the amount of OVII-bearing 
gas in the halo is sufficient to close the Galaxy's baryon census and (3) that a vast, $\sim 6$ kpc radius,  spherically-symmetric central region of the Milky Way above and 
below the 0.16 kpc thick plane, has either been emptied of hot gas (Model B) or the density of this gas within the cavity has a peculiar profile, increasing from the center up 
to a radius of $\sim 6$ kpc, and then decreasing with a typical halo density profile (Model A). 

The large value of $R_s$ in both the scenarios implied by Model A and Model B can be understood in terms of a radially expanding blast-wave or a shock-front generated in 
the center of the Galaxy and traveling outwards, so acting as a piston onto the ambient gas, and compressing the material at its passage, while pushing it (or a fraction of it) 
outwards. The central black hole of our Galaxy, could have played a fundamental role in this (e.g. $^{(12,13,14)}$), during a recent period of its activity. 
Faucher-Giguére \& Quataert (2012) study the property of galactic winds driven by active galactic nuclei, and show that energy-conserving outflows with initial velocity 
$v_{in} > 10000$ km s$^{-1}$, can move in the ambient medium producing shocked wind bubbles that expand at velocities of $v_s \simeq 1000$ km s$^{-1}$ into the host 
galaxy. 
If the observed OVII-bearing bubble in our Galaxy is tracing one of such shocks generated by our central supermassive black hole during a period of strong activity then, 
at a speed of 1000 km s$^{-1}$, the expanding shell would have taken 6 Myrs to reach its current radius of 6 kpc. Interestingly, $(6 \pm 2)$ Myr is also the age 
estimated for the two disks of young stars present in the central parsec of our Galaxy that are thought to be a relic of a gaseous accretion disk that provided fuel for 
AGN-like activity of our central black hole about 6 Myr ago $^{(15,16)}$. 

\section{Acknowledgements}
F.N. acknowledges support from the INAF-PRIN grant 1.05.01.98.10.
 
\section{References}
\end{document}